\documentclass[
    notitlepage,
    superscriptaddress,
    nofootinbib,
    amsmath,
    amssymb,
    amsfonts,
    aps,
    pra,
    showkeys,
    a4paper, 10pt
]{revtex4-2}

\usepackage{graphicx}%
\usepackage{subcaption}
\usepackage{dcolumn}%
\usepackage{bm}%
\usepackage[colorlinks=True]{hyperref}%
\usepackage{physics}
\usepackage[
activate={true,compatibility},
final,
tracking=true,
kerning=true,
nopatch=footnote]{microtype}

\usepackage{mathtools}

\usepackage[
text={6.7in,10in},centering,
]{geometry}

\usepackage{wrapfig}

\usepackage{color}

\DeclareRobustCommand{\rchi}{{\mathpalette\irchi\relax}}
\newcommand{\irchi}[2]{\raisebox{\depth}{$#1\chi$}}

\usepackage{booktabs}
\usepackage{tabularx}

\newcolumntype{M}{>{\centering}X}
\newcolumntype{Y}{>{\hsize=.22\textwidth\arraybackslash}X}
\newcolumntype{C}{>{\hsize=.22\textwidth\centering\arraybackslash}X}
\newcolumntype{R}{>{\hsize=.22\textwidth\raggedleft\arraybackslash}X}

\begin{document}

\title{Technical report on a quantum-inspired solver for simulating compressible flows}

\author{Raghavendra Dheeraj Peddinti}
\affiliation{Quantum Research Center, Technology Innovation Institute, Abu Dhabi, UAE}

\author{Stefano Pisoni}
\affiliation{Quantum Research Center, Technology Innovation Institute, Abu Dhabi, UAE}
\affiliation{Institute for Quantum-Inspired and Quantum Optimization, Hamburg University of Technology, Germany}

\author{Egor Tiunov}
\email{egor.tiunov@tii.ae}
\affiliation{Quantum Research Center, Technology Innovation Institute, Abu Dhabi, UAE}

\author{Alessandro Marini}
\affiliation{Propulsion and Space Research Center, Technology Innovation Institute, Abu Dhabi, UAE}

\author{Leandro Aolita}
\affiliation{Quantum Research Center, Technology Innovation Institute, Abu Dhabi, UAE}

\begin{abstract}
This document presents a quantum-inspired solver for 2D Euler equations, accepted at the final phase of the Airbus-BWM Group Quantum Computing Challenge (ABQCC) 2024. We tackle the case study of Quantum Solvers for Predictive Aeroacoustic and Aerodynamic modeling tasks.
We propose a tensor network based solver that scales polylogarithmically with the mesh size, in both runtime and memory. This provides a promising avenue for tackling the curse of dimensionality that plagues the direct numerical simulations in the field of computational fluid dynamics.
\end{abstract}


\maketitle
\tableofcontents
\clearpage
\section{Introduction}

While quantum computers promise a drastic advantage over classical ones, fault-tolerant machines are yet to be realized. Because of this, in our recent work~\cite{mpscfd2023}, we developed a quantum-inspired framework, based on tensor networks, for solving the incompressible Navier-Stokes equations. There, we developed a complete framework that simulates flows around immersed objects.
This framework fits perfectly with the problem posed. In fact, in Ref.~\cite{mpscfd2023}, we solve one of key challenges of tensor-network encoding the information about immersed objects.
In this submission, we propose a quantum-inspired solver for partial differential equations, as outlined in the forward track of Quantum Solvers. Our tensor network based solver scales polylogarithmically in the mesh size, in both runtime and memory. This provides a promising avenue for tackling the curse of dimensionality that plagues the direct numerical simulations in the field of computational fluid dynamics.

\subsection{Summary}
We propose to solve the \textit{2D Aerofoil Flow Prediction}\footnote{The 2D Aerofoil Flow Prediction problem is to simulate the flow around a NACA 0012 airfoil at transonic speeds.} problem with a quantum-inspired framework. The framework is based on tensor network methods, a powerful tool to simulate quantum dynamics.
In specific, we employ the well-known matrix product states (MPSs), also known as tensor train (TT).
These are efficient representations that mitigate the \textit{curse of dimensionality}, a major limitation common to both simulation of quantum physics and computational fluid dynamics.\\

In order to solve the specified partial differential equations (PDEs), our framework consists of several key components.
First, we represent all discretized variables with MPSs. The finite difference operators are represented by matrix product operators (MPOs).
Second, we encode the given geometry of the immersed object, e.g. the NACA0012 airfoil, into an MPS: \textit{MPS mask}. We then use these MPS masks to enforce the no-slip boundary condition on the velocity fields.
Here, given the analytical function describing the object's boundary, we find the approximate MPS mask using only few queries.
Third, we recast the existing classical numerical schemes to obtain the solutions within the MPS representation. Whenever the chosen numerical scheme results in solving a linear system, we tackle it with a DMRG-type algorithm.
Finally, we propose an \textit{MPS solution oracle} that retrieves the solution from the MPS encoding. We achieve this while bypassing the exponential overhead of converting the MPS back into the dense-vector representation.\\

The proposed framework is efficient and scalable, given that all algebraic operations are completely within the MPS representation, therefore never requiring the to deal with the large dense-vectors. The powerful memory compression is particularly relevant for industrial case studies, where numerical simulations are often limited by the huge number of parameters required by mesh-based solvers, for example in DNS simulations of turbulent flows.\\

One of the salient features is MPS mask generation where we can encode arbitrary geometries in MPSs, given only few queries to their functional description, using the TT-cross algorithm.
Another key feature is the MPS solution oracle, with two main operation modes: coarse-grained evaluation and pixel sampling. In particular, coarse-grained evaluation produces an averaged version of the solution. Instead, the latter generates unbiased samples according to a target property (vorticity, pressure, etc.). This is an ideal basis for developing aerodynamic design optimization software.\\

In this proposal, we implemented a steady state solver for the 2D Aerofoil Flow Prediction case study.  For the proposed solution, we analyze the asymptotic complexities of all the essential subroutines. We remark that, for a mesh size of $2^N$, the runtime scales linearly in $N$ and polynomially in bond dimension $\rchi$. Although our current implementation did not converge, we present empirical evidence that the steady state solution can be represented by MPSs of low bond dimension. This reinforces our belief that the MPS-based approach is valid for simulating shock-waves. Finally, we note that the stability and accuracy of the solver can be further improved as detailed in the Outlook. Additionally, in Appendix~\ref{app:AppendixB}, we present empirical evidence for the validity of the MPS-based approach in a 1D test case, comparing it with the known analytical solution.

\clearpage
\section{Detailed Explanation}\label{sec:detailed explanation}
\begin{figure*}[h]
     \includegraphics[width=\textwidth]{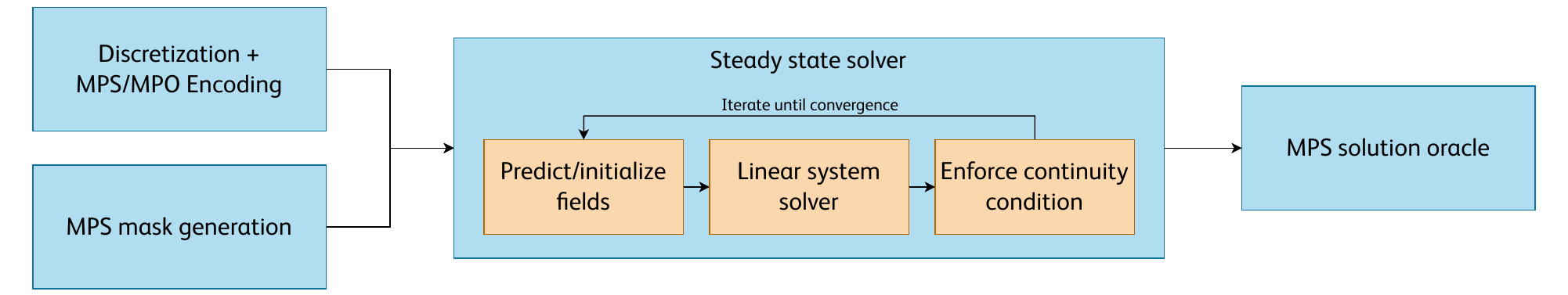}
     \caption{ \textbf{Schematics of the framework.} The figure shows the main components of the framework, explained in detail in the following subsections. In Sec.~\ref{subsection: MPS and MPO}, we detail discretization schemes and the choice of encoding for MPS/MPO representations. Next, in Sec.~\ref{subsection: Mask generation}, we explain the generation of approximate MPS masks. In Sec.~\ref{subsection:steady_state}, we present the numerical scheme to solve for the steady state solution. Finally, in Sec.~\ref{subsection:sampleoracle}, we explain the main features of the MPS solution oracle. In Sec.~\ref{sec:theoretical_validity}, we provide a detailed complexity analysis of the various components.
     }\label{fig:outline}
\end{figure*}

We solve the discretized Euler equations with an MPS framework. We summarize the framework in Fig.~\ref{fig:outline}. Our presentation is restricted to the problem statement of aerodynamic modeling, but all components of the framework can be straightforwardly generalized to the aero-acoustic problem as well.

\subsection{Discretization and MPS/MPO encoding}
\label{subsection: MPS and MPO}
We encode scalar functions -- such as the velocity components $v_x(\boldsymbol{x})$ and $v_y(\boldsymbol{x})$, or the pressure $p(\boldsymbol{x})$ -- into matrix-product states (MPSs)~\cite{schollwock2011density,orus2014practical,verstraete2008matrix}. 
We discretize the 2D domain into a mesh of $2^N$ points, specified by $N=N_x+N_y$ bits.
There, we represent a continuous function $v(x,y)$ by a vector of elements $v_{\boldsymbol i,\boldsymbol j} := v(x_{\boldsymbol i},y_{\boldsymbol j})$, where the bitstrings $(i_1,i_2, \hdots i_{Nx})$ and $(j_1,j_2, \hdots j_{Ny})$ are given by the binary representation of $\boldsymbol i$ and $\boldsymbol j$, respectively.
Then, we write each $v_{\boldsymbol i ,\boldsymbol j}$ as a product of $N$ matrices:
\begin{equation}
\label{eq:MPS}
    v_{\boldsymbol i,\boldsymbol j} = A_1^{(i_1)} A_2^{(i_2)}\dots A_{N_x}^{(i_{N_x})}\,B_1^{(j_1)} B_2^{(j_2)}\dots B_{N_y}^{(j_{N_y})}.
\end{equation}
This is the MPS representation. The indices $i_k$ and $j_k$ are referred to as \textit{physical indices} of the matrices $A_j$ and $B_j$, respectively. 
Note that, the first $N_x$ matrices correspond to $x_{\boldsymbol i}$ and the remaining $N_y$ ones to $y_{\boldsymbol j}$, as in~\cite{Greens_function_MPS,kiffner_jaksch2023tensor}. 
However, other arrangements are possible~\cite{Erika_vlasov_eq,Image_Compression,gourianov2022quantum}. For instance, one could consider an arrangement like $A_1B_1A_2B_2...A_NB_N$ that better highlights the structure of spatial scales.
The choice for the best encoding is problem-dependent and for the 2D airfoil simulation we choose the arrangement given in Eq.\eqref{eq:MPS}, while we compare both arrangements for the 3D turbulence case study that is presented in App. \ref{app:AppendixB}.

The bond dimension $\rchi$ is defined as the maximum dimension over all $2\,N$ matrices used.  
Importantly, the total number of parameters is at most $2\,N\,\rchi^2$. Hence, if $\rchi$ constant, the MPS provides an exponentially compressed representation of the $2^N$-dimensional vector. We remark that, given the binary discretization of the domain, the \textit{virtual indices} associated with the dimensionality of the matrices carry the correlations across different length scales. Their dimensionality can be kept fixed or dynamically adapted to the amount of inter-scale correlations. In this work, we choose to allow for a maximal $\rchi$.

In order to represent the spatial differential operators appearing in the PDEs, we adopt a central finite difference scheme. The resulting discretized operators are analytically mapped into the corresponding MPOs via known constructions~\cite{kazeev2012low}. Due to the simple structure of the discretized operators, their MPO counterparts have small $\rchi$'s ($<5$). This allow us to translate the actions of differential operators to efficient MPO-MPS contractions. The resulting MPS is then immediately truncated to the desired $\rchi$.

\subsection{MPS mask generation} \label{subsection: Mask generation}
If a rigid body is immersed in a fluid, the velocity field must be zero inside the object. Moreover, because of the inviscid nature of the flow, only the normal component of the velocity must vanish on the object boundaries.
The nullification of the velocity field inside the object is taken care by the masking technique, introduced in the context of quantum-inspired solvers in our recent work~\cite{mpscfd2023}.
To this end, we introduced the notion of \textit{approximate MPS masks}. These are MPSs encoding functions that approximate the target indicator function $\theta$ of the object -- i.e. the function equal to zero within the object and to one outside it -- up to arbitrary, tunable precision. Then, the nullification condition is imposed via simple element-wise multiplication between the vector field and the mask, which can be done efficiently in the MPS representation.
However, this procedure numerically results in enforcing all the components of the velocity field to be zero at the boundaries, instead of just nullifying the normal one. A more sophisticated implementation of the boundary conditions for inviscid flows might be the key to more accurate numerical simulations. This is a research direction that will be explored in the future.

\begin{wrapfigure}{r}{0.25\textwidth}
    \centering
    \includegraphics[width=0.25\textwidth]{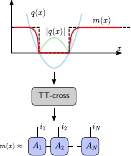}
    \caption{\textbf{MPS mask generation.} The indicator function $\theta(x)$ (dashed line) is approximated by the smooth function $m(x)$. Then the associated MPS is constructed via the TT-cross algorithm.}\label{fig:mask}
    \vspace{-0.5cm}
\end{wrapfigure}

As depicted in Fig.~\ref{fig:mask} for the one dimensional case, we build such MPSs from the smooth function $m(x)$:
\begin{equation} \label{eq:mask}
    m(\boldsymbol{\mathrm{x}}) \coloneqq 1 - e^{- \alpha \left(q(\boldsymbol{\mathrm{x}})
            + | q(\boldsymbol{\mathrm{x}}) |\right)},
\end{equation} with $\alpha > 0$. By construction, $m$ approximates $\theta$ (black dashed line in Fig.~\ref{fig:mask}) increasingly better for growing values of $\alpha$, and the nullified region corresponds to the inner part of the masked object. Moreover, $q({\mathrm{x}})$ is an auxiliary scalar function used to construct $m(x)$.

The next step is to obtain an MPS representation of $m(\boldsymbol{\mathrm{x}})$. If $m(\boldsymbol{\mathrm{x}})$ consists of elementary functions, the MPSs can be constructed analytically~\cite{Oseledets_2012_real_functions,garcia_ripoll2021quantum}. However, in general, one must resort to numerical approximations. In this work, we use the standard TT-cross algorithm~\cite{OSELEDETS201070TTcross} to find an MPS approximation of $m(\boldsymbol{\mathrm{x}})$. This extends the standard cross interpolation algorithm to MPSs that requires only $\mathcal{O}(N\rchi^2)$ evaluations of the function to obtain the MPS approximation up to a controllable $l_2$ error $\epsilon$, where $\rchi$ is the bond dimension of the approximate MPS~\cite{OSELEDETS201070TTcross}. This querying procedure learns the MPS representation in a fashion similar to the geometry-informed neural networks~\cite{li2023geometryinformed}.

Finally, as demonstrated in~\cite{mpscfd2023}, we remark that the need for a smooth approximation $m$ to the indicator function $\theta$ stems from the empirical observation that direct TT-cross approximations to $\theta$ produce significantly worse masks than via the intermediate, tunable, approximation $m$.

\subsection{Steady state solver} \label{subsection:steady_state}
The 2D Aerofoil Flow Prediction problem is to simulate the flow around a NACA 0012 airfoil at transonic speeds. At such speeds, the compressibility of air causes shock waves and the aim is to simulate this behavior of compressible flows. In this study, we simplify the problem to solve for the steady state behavior of inviscid fluids. The dynamics of compressible and inviscid fluids are governed by the Euler equations, shown in Eqs.~\ref{eq:Euler_eq}.

\begin{subequations} \label{eq:Euler_eq}
\begin{align} 
    \pdv{\rho}{t} + \div{(\rho \Vec{u})} &= 0 \label{eq:Euler_eq_1} \\
    \pdv{\Vec{u}}{t} + \div{(\rho \Vec{u} \Vec{u}^T)} + \grad{p} &= 0 \label{eq:Euler_eq_2} \\
    \pdv{E}{t} + \div{(\Vec{u}(E + p))} &= 0 , \label{eq:Euler_eq_3}
\end{align}
\end{subequations}

where $p$ is the static pressure, $\rho$ is density of the fluid. The total energy is given by $E = \frac{p}{\gamma - 1} + \frac{1}{2} \rho u^2$ and $\gamma$ is the ratio of specific heats ($\gamma = 1.4$ for air). 

Now, we look at the steady state formulation of this problem by setting all the temporal derivatives to zero. The resulting equations in their conservative form for the two dimensional case are presented in Eqs.\ref{eq:steady_euler}.
\begin{subequations} \label{eq:steady_euler}
\begin{align} 
    \pdv{(\rho u)}{x} + \pdv{(\rho v)}{y} &= 0 \label{eq:continuity} \\
    \pdv{(\rho u^2 + p)}{x} + \pdv{(\rho uv)}{y} &= 0 \label{eq:momentum_x} \\
    \pdv{(\rho uv)}{x} + \pdv{(\rho v^2 + p)}{y} &= 0 \label{eq:momentum_y} \\
    \pdv{((E+p) u)}{x} + \pdv{((E+p) v)}{y} &= 0  \label{eq:energy_eq}
\end{align}
\end{subequations}

We use the second-order central finite difference scheme to discretize the equations. We encode the difference operators into MPOs~\cite{kazeev2012low}, along with the corresponding boundary conditions. However, due to discontinuous nature of the solution, the central difference schemes become unstable. This instability arises from the high-frequency oscillations at the discontinuity known as the Gibbs phenomena~\cite{Shu1998}. However, several stabilization techniques are available, ranging from artificial dissipation terms~\cite{jst1981} to essentially non-oscillatory (ENO) upwinding schemes~\cite{Shu1998}.

In this work, we use the simplest artificial dissipation term, given by a second-order term with a constant coefficient $\mu$. Extension to higher-order terms and adaptive dissipation coefficients is straightforward at the expense of computational overhead. We also linearize the non-linear equations by starting with an initial guess for the variables $\rho^{(0)}, (\rho u)^{(0)}, (\rho v)^{(0)},$ and $(E + p)^{(0)}$ and updating them through several iterations until convergence. After the linearization of Eqs.~\ref{eq:steady_euler} and addition of artificial dissipative terms, the resulting coupled linear system, given in Eqs.~\ref{eq:linear_sys}, is solved iteratively using a DMRG-based linear system solver~\cite{oseledets2012solution}.

\begin{equation} \label{eq:linear_sys}
(\Hat{O}_{adv} + \Hat{O}_{diss})\begin{pmatrix}
    \rho^{(i)}\\
    (\rho u)^{(i)} \\
    (\rho v)^{(i)}\\
    (E + p)^{(i)}
\end{pmatrix}
= 
-\begin{pmatrix}
    \rho^{(i-1)}\\
    (\rho u)^{(i-1)} \\
    (\rho v)^{(i-1)}\\
    (E + p)^{(i-1)}
\end{pmatrix}(\div \Vec{u}^{(i-1)})
-\begin{pmatrix}
    0\\
    \partial_x p^{(i-1)} \\
    \partial_y p^{(i-1)} \\
    0
\end{pmatrix},
\end{equation} 
where $\Hat{O}_{adv} = u^{(i-1)} \partial_x + v^{(i-1)} \partial_y$ and
$\Hat{O}_{diss} = \mu (\partial_{xx} + \partial_{yy})$.

Here, we have multiple strategies to solve the coupled linear system. First, one can segregate each of the equation and individually update the corresponding variables in chosen order. Secondly, one can build a larger linear system with all of components solved simultaneously. We choose the first segregated approach unless mentioned otherwise.  A combined approach is also worth exploring.

Additionally, at the end of each iteration, we enforce the continuity equation (Eq.~\ref{eq:continuity}) by projecting the flux components $\rho u$ and $\rho v$ on to their divergence free manifold. To this end, we utilize the DMRG-based Poisson solver from our previous work~\cite{mpscfd2023}, where the Chorin's projection method was used to enforce the divergence-free condition.

\subsection{MPS solution oracle}\label{subsection:sampleoracle}

\begin{figure}
    \centering
    \includegraphics[width=0.7\textwidth]{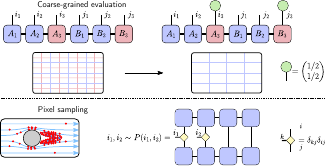}
    \caption{\textbf{Schematic of the solution oracle.}
    The figure depicts the two proposed methods to extract the main features of the solution from the MPS itself: coarse-grained evaluation and pixel sampling. Both of them avoid mapping back the MPS to the dense-vector representation, which could eliminate the potential speed-ups gained in solving the problem.
    }\label{fig:solution_oracle}
\end{figure}

Our solver outputs the solution in the MPS form, requiring up to exponentially fewer parameters than the dense-vector representation.
This raises the question of how to retrieve the solution without mapping the MPS to the dense vector, which can eliminate potential speed-ups gained in solving the problem. We propose two methods to extract the solution directly from the MPS: {\it coarse-grained evaluation} and {\it pixel sampling}.

Coarse-grained evaluation (Fig.~\ref{fig:solution_oracle}, top) produces an MPS with fewer spatial scales, encoding an averaged version of the solution. 
The averaging is computed by efficiently contracting the single MPS tensors, encoding the spatial scales to be averaged, with local constant tensors (in light-green in the figure). The resulting smaller MPS can then be mapped to the dense vector representation by standard contractions.
Clearly, the finest-scales details are lost in this operation mode. 
A convenient, complementary alternative is pixel sampling.

Pixel sampling (Fig.~\ref{fig:solution_oracle}, bottom) is an exact sampling method that generates random points $(x_{\boldsymbol{\mathrm{i}}},y_{\boldsymbol{\mathrm{j}}})$ chosen with probability $P({\boldsymbol{\mathrm{i}}},{\boldsymbol{\mathrm{j}}})=|v_{\boldsymbol{\mathrm{i}},\boldsymbol{\mathrm{j}}}|^2/\|{\boldsymbol{\mathrm{v}}}\|_2$, where $\|{\boldsymbol{\mathrm{v}}}\|_2$ is the $l_2$-norm of ${\boldsymbol{\mathrm{v}}}$.
Next, one can efficiently query the MPS to assess the encoded solution at the sampled points.
This reproduces the measurement statistics of a quantum state that encodes $\boldsymbol{\mathrm{v}}$ in its amplitudes. We note that related sampling methods have been used for training mesh-free neural-network solvers \cite{SIRIGNANO20181339} (interestingly, the method developed in this reference was a subject of study in the previous Q challenge by Airbus in 2019). 
Our approach relies on standard techniques for $l_2$-norm sampling physical-index values from an MPS~\cite{orus2014practical,schollwock2011density}. 
Importantly, the technique applies to any MPS~\cite{ferris2012perfect,han2018unsupervised}. For instance, one can obtain the MPS encoding the vorticity field from that of the velocity. 
\begin{figure}[htbp]
    \centering
    \begin{minipage}[b]{0.47\textwidth}
        \centering
        \includegraphics[width=\textwidth]{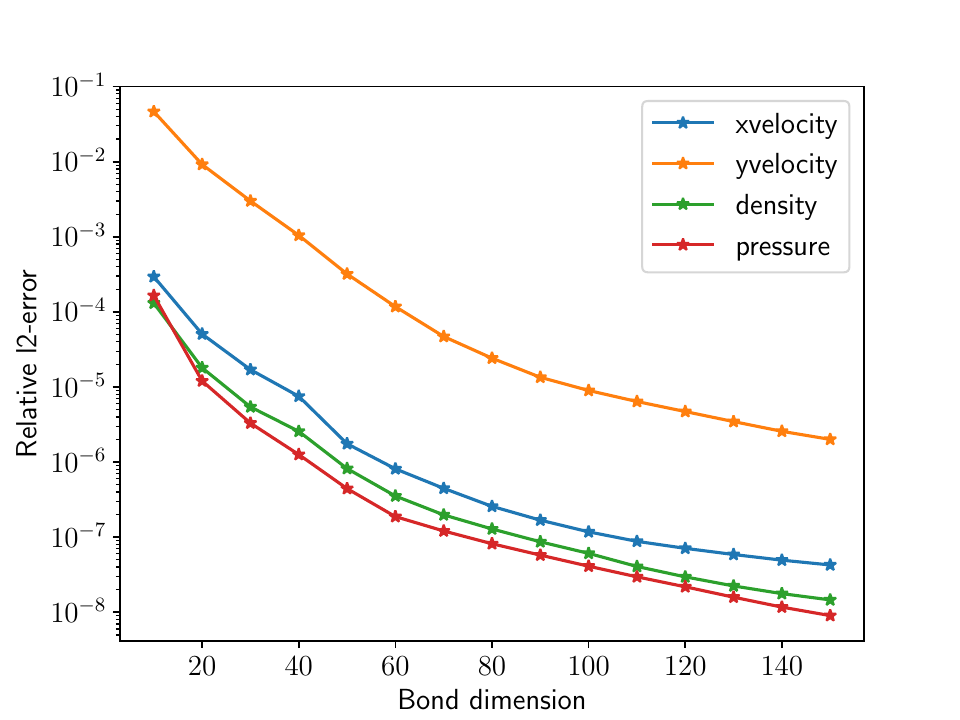}
        \caption{\textbf{Compression error of the steady state solution.} Relative $l_2$-error resulting from encoding the vectors into MPS of varying bond dimension. The vectors correspond to the various fields of steady state solution from Ansys Fluent.}\label{fig:l2_error}
    \end{minipage}
    \hfill
    \begin{minipage}[b]{0.47\textwidth}
        \centering
        \begin{tabularx}{\textwidth}{@{}XCC@{}}
             \toprule
             \toprule
             \textbf{Algorithmic task} & \textbf{Complexity} & \textbf{Status}\\
             \toprule
             Mask generation & $\mathcal{O}(N\,\rchi^3)$ & \texttt{offline} \\
             \midrule
             Mask application & $\mathcal{O}(N\rchi^4)$ & \texttt{online} \\
             \midrule
             Linear system solver & $\mathcal{O}(N\rchi^4)$ & \texttt{online} \\
             \midrule
             Coarse-grained evaluation & $\mathcal{O}(N\,\rchi^3)$ & \texttt{offline} \\
             \midrule
             Pixel sampling (per pixel) & $\mathcal{O}(\rchi^3)$ & \texttt{offline} \\
             \bottomrule
             \bottomrule
        \end{tabularx}
        \captionof{table}{\textbf{Time complexities.} The first column shows the main subroutines. Their corresponding asymptotic worst-case time complexity scaling and their status during the time stepping scheme are shown in the second and third columns, respectively. We refer by \texttt{offline} to the tasks performed outside the steady state solver loop.}\label{table:complexity}
    \end{minipage}
\end{figure}

\subsection{Theoretical validity and scaling} \label{sec:theoretical_validity}
Here, we present the theoretical validity of the numerical scheme and the MPS framework, along with some empirical evidence. We also provide a thorough complexity analysis of the solver.

Firstly, in Ref.~\cite{swanson1997comparison}, the authors study the equivalence between the artificial dissipation and upwind schemes. For the linearized equations, the second-order dissipation operator can be derived starting from a first-order upwind scheme. For the choice of artificial dissipation $\mu = h$ where h is the grid size, $\mu$ vanishes under mesh refinement as $h \xrightarrow{} 0$. Thus, with very refined meshes, we recover the inviscid form of the equations. This is particularly advantageous for the MPS framework as it can represent exponentially refined meshes efficiently.

Next, we present theoretical and empirical arguments that point to favorable bond dimensions. In Refs.~\cite{gourianov2022quantum} and~\cite{lindsey2023multiscale}, the authors provide analytical constructions on encoding 1D discontinuities in MPS representations with low bond dimensions, with a specified accuracy. Although, such analytical arguments are not available for 2D discontinuities, we empirically observe that the bond dimensions stay low. Specifically, in Fig.~\ref{fig:l2_error}, we notice the steady state solution of the Euler equations around the NACA 0012 airfoil can be compressed with MPS encoding with relatively high accuracy.

Next, we present the asymptotic complexity analysis of the solver and its components. Table~\ref{table:complexity} shows the theoretical scaling per subroutine. Tasks performed outside of the steady state solver subroutine are tagged as \texttt{offline}, as they are performed independently.
The complexity of the mask generation is due to the TT-cross approximation~\cite{OSELEDETS201070TTcross}. The mask application step consists of multiplication of MPSs along with the truncation of bond dimension. We utilize the Hadamard-avoiding TT compression algorithm, with complexity $\mathcal{O}(N\rchi^4)$~\cite{Saibaba_2023,sun2024hatt}. The linear system solver is dominated by a sub-task involving $N$ linear systems of size $4\rchi^2\times 4\rchi^2$, which are solved approximately via iterative approaches, with complexity $\mathcal{O}(N\rchi^4)$~\cite{gourianov2022quantum}. The complexity of other tasks are dominated by their corresponding tensor-network contractions~\cite{schollwock2011density,ferris2012perfect}.

\section{Implementation}
We simulate the flow around a NACA0012 airfoil at transonic Mach numbers. We analyze and compare the results obtained with the MPS solver against Ansys Fluent, a standard CFD software. Additionally, for the 1D Euler equations, we present a benchmark of the MPS solver against the analytical solution in App.~\ref{sec:1d_case}. 
For the case of 2D flow prediction around the NACA 0012 airfoil, we solve the Euler equations for a grid size of $2048\times1024$, using the stacked encoding for the MPS. For the boundary conditions of the domain, we enforce an inlet condition on the left and an outlet condition on the right end of the domain. The top and bottom boundaries are treated periodically. The Ansys Fluent simulations were run on a non-uniform grid of roughly 2.4 million cells. They use a second order upwind scheme and finite volume discretization to solve the Euler equations.
\begin{figure}[htbp]
    \begin{subfigure}[b]{0.45\linewidth}
        \centering
        \includegraphics[width=\linewidth]{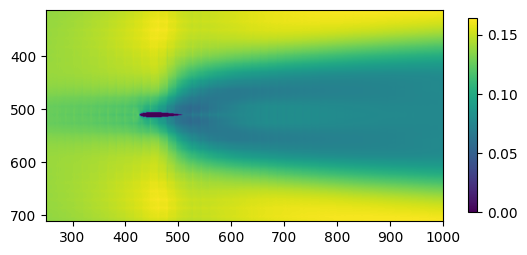} %
        \caption{Pressure field of MPS solver}
        \label{fig:mps_pressure}
    \end{subfigure}
    \hfill
    \begin{subfigure}[b]{0.45\linewidth}
        \centering
        \includegraphics[width=\linewidth]{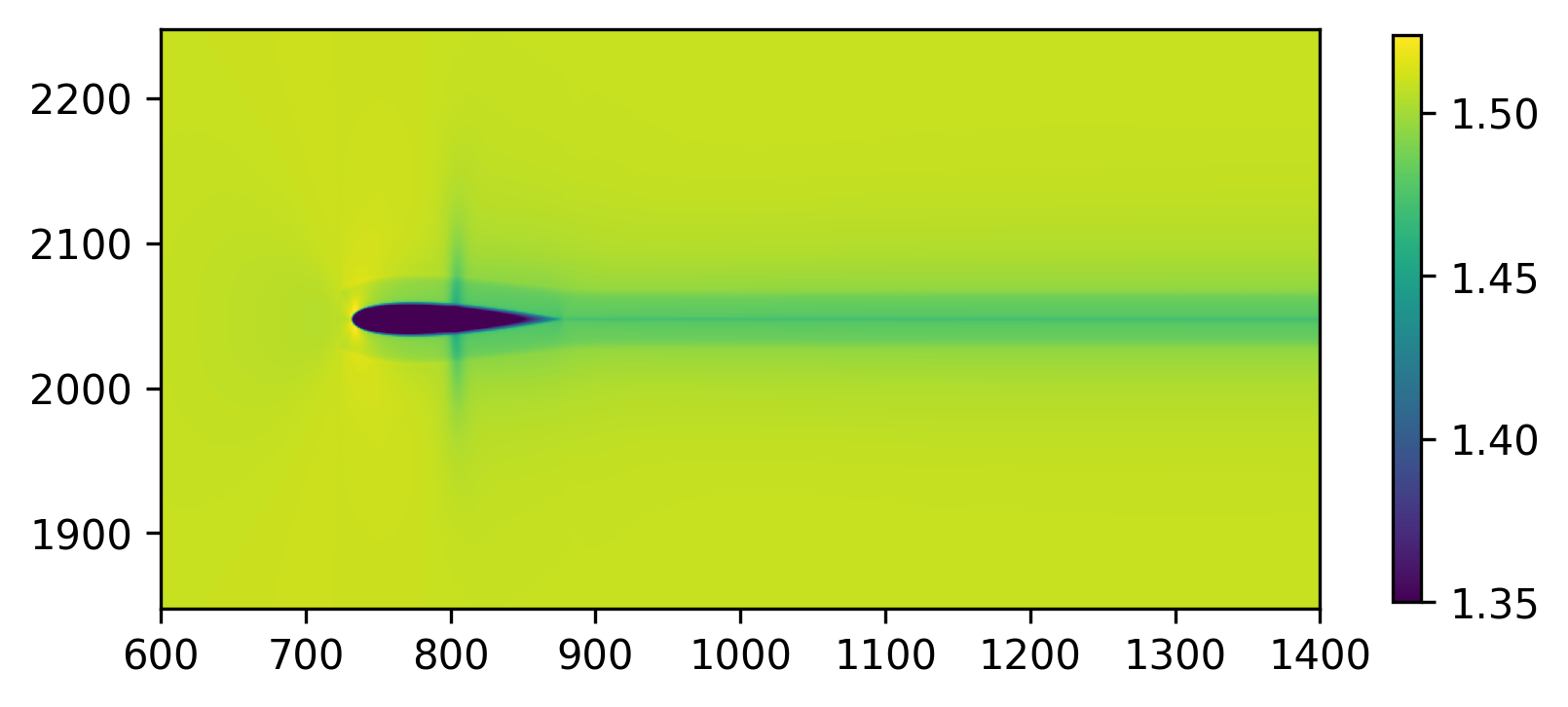} %
        \caption{Pressure field of Ansys Fluent}
        \label{fig:ansys_pressure}
    \end{subfigure}
    \begin{subfigure}[b]{0.45\linewidth}
        \centering
        \includegraphics[width=\linewidth]{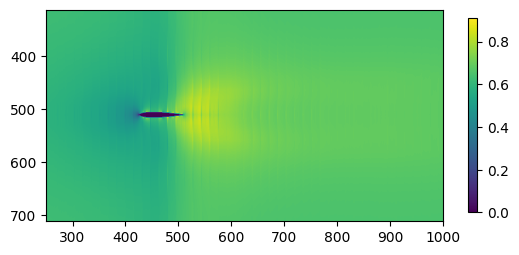} %
        \caption{Velocity magnitude of MPS solver}
        \label{fig:mps_velmag}
    \end{subfigure}
    \hfill
    \begin{subfigure}[b]{0.45\linewidth}
        \centering
        \includegraphics[width=\linewidth]{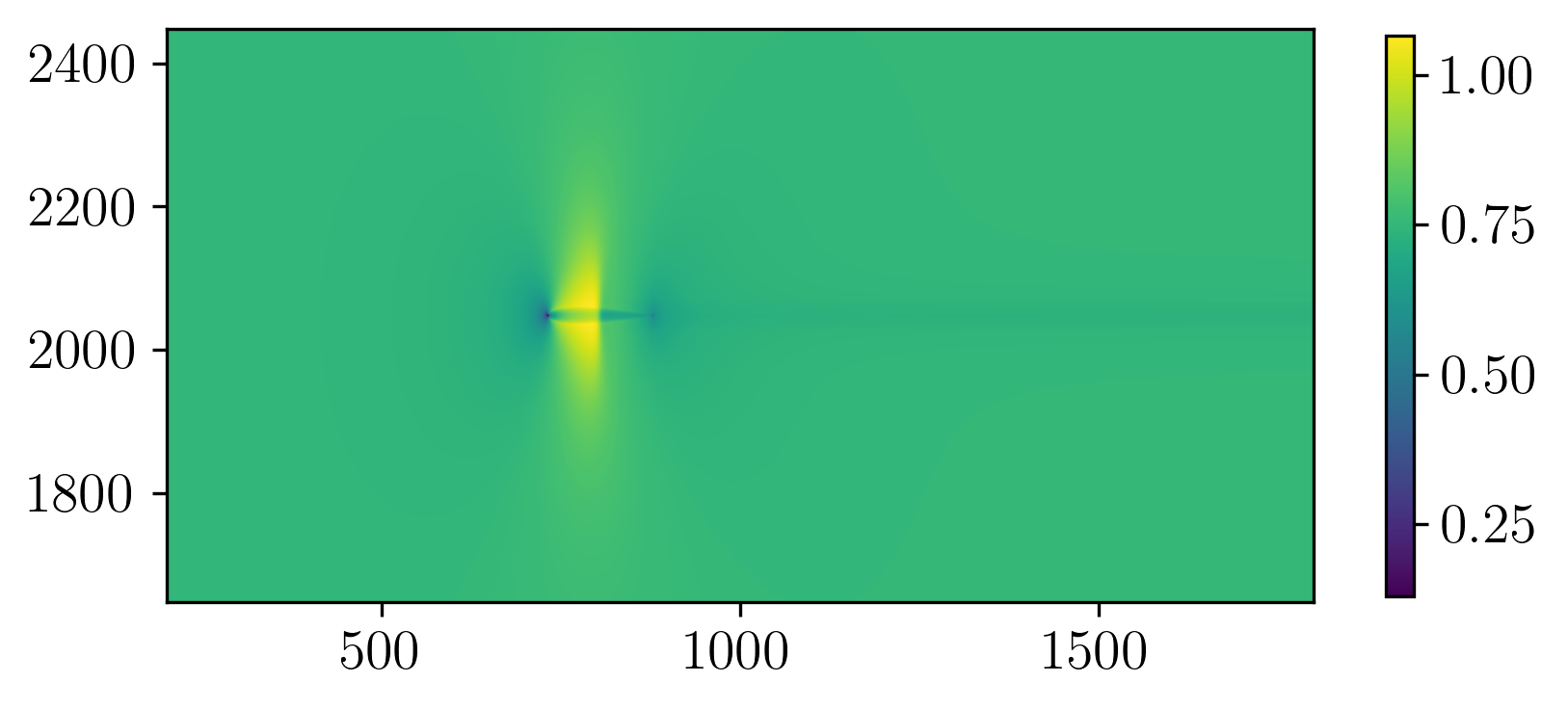} %
        \caption{Velocity magnitude of Ansys Fluent}
        \label{fig:ansys_velmag}
    \end{subfigure}
    
    \caption{\textbf{Simulation results of MPS framework and Ansys Fluent.} Here, we present the pressure and velocity magnitudes from the MPS solver and Ansys Fluent. Since the MPS solver failed to converge, we present results from the last stable iteration. For the 21-bit MPS encoding, we allow for a maximum bond dimension of $\rchi=100$ during the simulation. The MPS mask for the NACA0012 airfoil has a bond dimension of 18. In panels (a) and (b) we plot the total pressure, $p+\frac{1}{2}\rho u^2+\frac{1}{2}\rho v^2$, divided by $\rho_0 u_0^2$ where $\rho_0=1.225$ kg/m$^3$ and $u_0=272.6$ m/s are the freestream values. In panels (c) and (d), we present the velocity magnitude of the fluid, $\sqrt{u^2 + v^2}$, divided by the speed of sound $c=340.2$ m/s.}
    \label{fig:simulation_results}
\end{figure}

We simulated the steady state solution using the described framework for several iterations and observed a lack of convergence. In Fig.~\ref{fig:simulation_results}, we present the solutions from the last stable iteration of the MPS solver and compare them against the converged solutions of Ansys Fluent. It can be seen that the MPS solutions do not match those of Ansys Fluent. However, one can observe general characteristics of the flow which point to the formation of a shock-wave. We note that the efforts are underway to improve both the accuracy and stability of the MPS framework. In the Outlook, we point out possible sources of instability and potential remedies.

\begin{figure}[h]
    \centering
    \begin{subfigure}{0.45\linewidth}
        \centering
        \includegraphics[width=\linewidth]{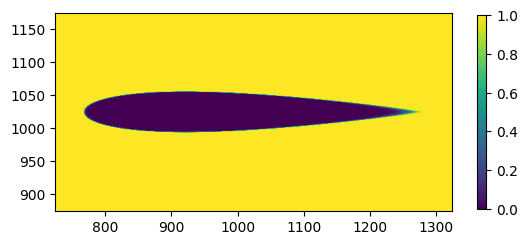} 
        \label{fig:mask_smooth}
    \end{subfigure}%
    \hfill
    \begin{subfigure}{0.45\linewidth}
        \centering
        \includegraphics[width=\linewidth]{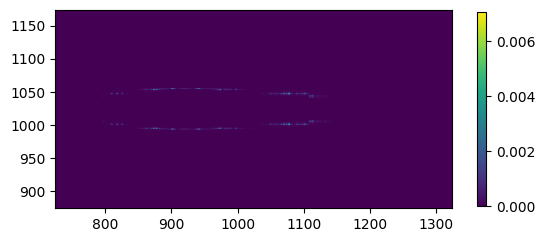} 
        \label{fig:mps_mask_error}
    \end{subfigure}
    \caption{\textbf{Reconstruction error of the MPS mask.} Here, we present (left) the smooth approximation $m(\boldsymbol{\mathrm{x}})$ of the NACA0012 airfoil and (right) the reconstruction error of the MPS mask, generated using $m(\boldsymbol{\mathrm{x}})$. We use a sharpness parameter of $\alpha=10^4$ to approximate the geometry. The domain of 16 m $\times$ 8 m with the chord length 1 m is discretized using $8192\times4096$ points. The resulting 25-bit MPS mask has bond dimension $\rchi=96$.}
    \label{fig:naca0012_mask}
\end{figure}

\textbf{Geometry and MPS mask:} For the NACA 0012 airfoil geometry, we observe that the smooth approximation $m(\boldsymbol{\mathrm{x}})$ results in a low bond dimension MPS mask. In Fig.~\ref{fig:naca0012_mask}, we present the reconstruction error of the MPS mask for one such instance. We can observe that the trailing edge of the wing is  slightly softened due to the smooth approximation.

\section{Conclusion $\&$ Outlook}

In conclusion, our proposal offers the following advantages:
\begin{itemize}
    \item Our proposed algorithm is inspired by quantum techniques but is fully classical in nature. That is, it does not require quantum hardware for its execution, offering a strategic advantage before fault-tolerant quantum computers appear. Interestingly, however, in the future, subroutines (linear system solver, e.g.) can naturally upgraded to quantum version as hardware advances.
    \item The framework is fully versatile, straightforwardly adaptable to other types of PDEs. In particular, one can extend the current implementation to also tackle the aero-acoustic case study.
    \item For a mesh of size $2^N$, both memory and runtime scale linearly in $N$ and polynomially in the bond dimension $\rchi$.
    \item Our solver supersedes previous tensor-network solvers in that it incorporates immersed objects of diverse geometries, with non-trivial boundary conditions. This is a crucial development by our team, which enables us to tackle realistic industry problems. 
    \item Another important feature of our solver is that can efficiently retrieve the solution directly from the compressed MPS encoding, i.e. without passing through the expensive dense-vector representation.

\end{itemize}

As potential improvements of our approach, it is important to mention the following points:
\begin{itemize}
    \item The most sensitive component of our algorithmic pipeline is the required bond dimension , and how to keep it under control during the integration.  In particular, the bond dimension of the MPS mask of an immersed object could be very large for non-trivial geometries. However, for geometries such as the airfoil, we consistently find efficient MPS approximations. 
    \item In turn, even if the mask bond dimension is low, it is in general unknown if $\rchi$ can be kept under control (by sequential truncation) without introducing significant approximation errors during the time evolution. 
    \item It was observed that the current implementation of the iterative steady state solver did not converge. We believe this can be fixed with a better treatment of the object's boundary conditions which satisfy the properties of inviscid flows, as noted in Sec.~\ref{subsection: Mask generation}.
    \item Additionally, standard density-based solvers use a coupled linear system solver for the continuity (Eq.~\ref{eq:continuity}) and the momentum (Eqs.~\ref{eq:momentum_x} and~\ref{eq:momentum_y}) equations. However, we use a segregated approach in our implementation which could lead to numerical instabilities.
\end{itemize}

Finally, on a more general perspective, our proposed solver scheme offers also several promising spin-off directions. In particular, the extension to the 3D case will enable probing the solver in truly turbulent regimes.
This will also require in-depth studies of the dependence of the bond dimension with time evolution. We anticipate that further optimization of the tensor-network Ansatz will play a crucial role for that, as well as extensions to non-uniform meshes. Another possibility is the exploration of tensor networks with other natural function bases for turbulent phenomena, such as Fourier~\cite{li2021fourier} or wavelets~\cite{mariefarge_wavelets}. In turn, an interesting opportunity is the application of our framework to other partial differential equations \cite{Greens_function_MPS,truong2023tensor,Lubasch_multigrid_renorm}. This may be combined with extensions to finite elements or finite volumes, which can  also be formulated with tensor networks~\cite{kornev2023chemicalmixer}.\\

This challenge has been instrumental in advancing our quantum-inspired solver towards relevant industrial case studies. Firstly, the proposed approach could greatly reduce the runtime and memory requirements for CFD simulations. Additionally, it could be the basis for developing other industrial tools. For instance, pixel sampling is ideal for Monte-Carlo simulations, relevant for aerodynamics design optimizations as well as for the training of mesh-free neural networks~\cite{SIRIGNANO20181339}. Finally, our work opens a playground with potential to build more efficient solvers of real-life fluid dynamics problems as well as other high-dimensional partial differential equation systems. This can have a major impact on industries that benefit from fluid dynamics simulations, in particular the automotive and aviation sectors.

\bibliography{apssamp}

\clearpage
\appendix

\section{Additional results for 1D case}\label{app:AppendixB}
In this appendix we provide additional evidence to support the validity of our approach. Here, we discuss how a shock-wave in 1D can be captured via an MPS.

\subsection{1D Euler equations} \label{sec:1d_case}

We study the Sod shock tube problem, a canonical test case for CFD solvers. They are described by the one-dimensional Euler equations, i.e., $\Vec{u} = u$. The initial conditions of the system are shown in Eq.~\ref{eq:shock_tube_ic} where the subscripts $L$ and $R$ refer to the left and right halves of the domain, respectively.

\begin{equation} \label{eq:shock_tube_ic}
    \begin{pmatrix}
    \rho_L \\
    p_L \\
    u_L
\end{pmatrix}
=
\begin{pmatrix}
    1.0 \\
    1.0 \\
    0.0
\end{pmatrix}, \quad
\begin{pmatrix}
    \rho_R \\
    p_R \\
    u_R
\end{pmatrix}
=
\begin{pmatrix}
    0.125 \\
    0.1 \\
    0.0
\end{pmatrix}
\end{equation}

\begin{figure}[h]
    \centering
    \includegraphics[width=\linewidth]{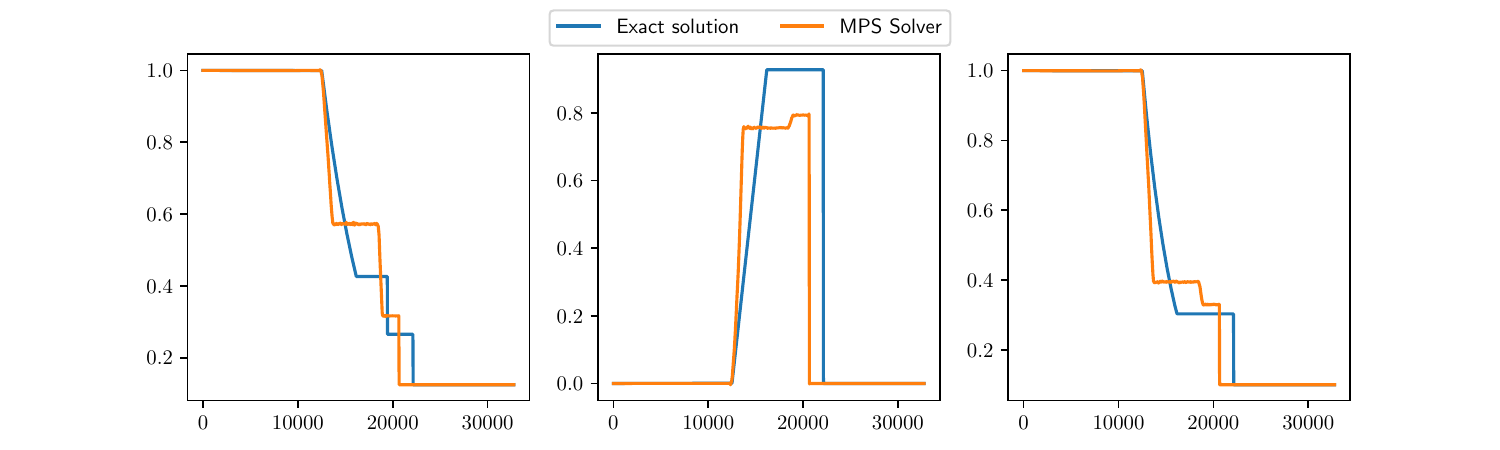}
    \caption{\textbf{Benchmark of MPS framework against analytical solution.} We compare the solutions of the MPS solver against the exact solutions after $t=0.1s$ of evolution. The 1D domain is discretized with $2^{15} = 32768$ points. (left) Density of the fluid is shown, where the MPS representation has 14960 parameters in total, resulting in $55\%$ reduction. (center) Velocity of the flow is encoded in a MPS with just 410 parameters, a remarkable $98.7\%$ compression. (right) The pressure of the fluid is shown, and is encoded in an MPS with 17304 parameters, with $47\%$ compression.}
    \label{fig:sod_shock_tube}
\end{figure}

\end{document}